\documentclass[a4paper]{article}

\usepackage{amsmath,amsfonts,graphicx,epsfig,mathrsfs}
\newcommand{\R}{{\mathbb{R}}}

\newcommand{\C}{{\mathbb{C}}}

\newcommand{\CP}{{\mathbb{C}}{{P}}}

\newcommand{\beq}{\begin{equation}} 
\newcommand{\eeq}{\end{equation}} 
\newcommand{\bea}{\begin{eqnarray}} 
\newcommand{\eea}{\end{eqnarray}} 
\newcommand{\ra}{\rightarrow}

\newcommand{\cd}{\partial}

\newcommand{\M}{{\sf M}} 
\newcommand{\eps}{{\varepsilon}}

\newcommand{\av}{{\bf A}} 
\newcommand{\bv}{{\bf B}} 
\newcommand{\avc}{{\sf A}} 
\newcommand{\bvc}{{\sf B}} 
\newcommand{\Jc}{{\sf J}} 
\newcommand{\nv}{{\bf n}} 
\newcommand{\mv}{{\bf m}} 
\newcommand{\sv}{{\bf S}}

\newcommand{\zv}{{\bf 0}} 
 
\newcommand{\DD}{\mathscr{D}}

\begin{document}

\title{Sigma models on curved space and
bubble refraction in doped antiferromagnets}

\author{J.M. Speight\\
School of Mathematics, University of Leeds, Leeds LS2 9JT, UK
}

\date{}
\maketitle

\begin{abstract}
The dynamics of bubble solitons in two-dimensional
isotropic antiferromagnets, 
inhomogeneously doped so
that the exchange integral $J$ becomes position dependent, is studied.
In the usual continuum approximation, the system reduces to
a nonlinear sigma model on a spacetime whose geometry depends on
$J(x)$. It is shown, both within the geodesic approximation, and
by appealing to field theoretic conservation laws, that a bubble
incident on a domain-wall inhomogeneity undergoes refraction in accordance
with Snell's law, $J^{-1}$ being
identified with the refractive index, and that sufficiently oblique
impacts result in total internal reflexion. Possible applications of
this phenomenon to the construction of bubble lenses and bubble guides
(in analogy with fibre-optic cables) are considered.

\end{abstract}

\section{Introduction}

Topological solitons have important applications in many branches of physics,
especially condensed matter and high energy physics. It often happens that
the same soliton can be given completely different physical interpretations
in these two disciplines, which has led in recent years to a very fruitful
exchange of ideas and techniques. One of the most surprising
and least exploited of these soliton correspondences is that between the
isotropic Heisenberg antiferromagnet and the relativistic $O(3)$ sigma
model. In this paper we will exploit this correspondence to show, using
techniques from particle theory and differential geometry, that the
trajectories of
bubble-like solitons in inhomogeneously doped two-dimensional
antiferromagnets experience refraction
in exact analogy with Snell's law of refraction in geometric optics. 

The isotropic Heisenberg antiferromagnet consists of an infinite square
lattice of classical spin variables $\sv_{ij}\in\R^3$, with 
$|\sv_{ij}|^2=s^2$, evolving according to the law
\beq\label{dag} 
\frac{d\sv_{ij}}{d\tau}=-\sv_{ij}\times\frac{\cd H}{\cd \sv_{ij}},
\qquad
H:=\sum_{i,j}J\left[2s^2+\sv_{ij}\cdot(\sv_{i,j+1}+\sv_{i+1,j})\right]
\eeq
where $\tau$ is time and $J>0$ is a constant called the exchange integral.
Note that since $J>0$, $H$ is minimized when each spin anti-aligns with
its nearest neighbours. It is a curious fact that this dynamical system,
first order in time and with an obvious preferred reference frame,
can be described, in the continuum limit, by a Lorentz invariant PDE
which is second order in time, namely
\beq
\label{*}
\nv\times \DD\nv =\zv,\qquad
\DD\nv:=\frac{\cd^{2}\nv}{\cd t^2}-J^2\left(\frac{\cd^2\nv}{\cd x^2}+
\frac{\cd^2\nv}{\cd y^2}\right)
\eeq
where $\nv:\R^{2+1}\ra\R^3$ has $|\nv|^2=1$, and $t$ is a
rescaled time variable. Note that (\ref{*}) holds 
if and only if $\DD\nv$ is parallel to $\nv$, and hence 
\beq
\DD\nv-(\nv\cdot\DD\nv)\nv=\zv,
\eeq
which is the more familar form of the $O(3)$ sigma model (or wave map)
equation. The connexion between $\nv$ and $\sv_{ij}$ is subtle, and
will be explained in the next section. Roughly speaking, one should
imagine the field $\nv$ sampled on a square lattice, partitioned into
black and white sublattices, chess-board fashion. The white spins evolve
as $s\nv(t)$, the black as $-s\nv(t)$.

Imagine now that the antiferromagnet is inhomogeneously doped, so that the 
exchange integral becomes a function of position, $J(x,y)$. 
We will argue that the spin dynamics is now described by a field
$\nv$ obeying
\beq\label{**}
\nv\times\left[\frac{\cd^{2}\nv}{\cd t^2}-J(x,y)^2
\left(\frac{\cd^2\nv}{\cd x^2}+
\frac{\cd^2\nv}{\cd y^2}\right)\right]=\zv,
\eeq
which is the $O(3)$ sigma model equation on spacetime $\R\times\R^2$ equipped
with an inhomogeneous Lorentzian metric
\beq
\eta=dt^2-\frac{1}{J(x,y)^2}(dx^2+dy^2).
\eeq
Note that the static field equation is independent of $J$, so the usual 
Belavin-Polyakov lumps of the homogeneous system \cite{belpol}, 
which we shall call
bubbles in this context (in analogy with magnetic bubbles in ferromagnets), 
carry over unchanged to the doped system. 
In this paper we make a detailed study of the dynamics of a single bubble
in various inhomogeneous antiferromagnets, primarily within the geodesic
approximation of Manton \cite{man}. The main calculational task in this 
approach is to compute the metric on the one-bubble moduli space. This
metric does depend on $J(x,y)$, and we obtain exact expressions for
it for several interesting choices of this function. 

 Our main conclusion is that bubble
trajectories experience refraction, in exact analogy with geometric
optics, $J(x,y)^{-1}$ playing the role of the refractive index of the medium.
The incident and exit angles of a bubble crossing a domain-wall 
$J$-inhomogeneity, for example, are related by Snell's law, and total
internal reflexion occurs if the impact is sufficiently oblique. This
key observation is rederived without reference to the geodesic approximation
by appealing to field theoretic conservation laws. We go on to study
simple prototype bubble ``lenses'' and bubble guides, which use total
internal reflexion to guide bubble trajectories just as fibre optic cables
guide light beams. 

The rest of the paper is structured as follows. In section \ref{deriv} we
derive the PDE (\ref{**}) from (\ref{dag}). In section \ref{geoapp}
the geodesic approximation
to the dynamics of a single bubble is set up. The interaction of a
bubble with a domain wall is studied in \ref{domwal} and Snell's law
of refraction obtained. Sections \ref{cirlen} and \ref{bubgui} consider
simple lenses and bubble guides respectively, while \ref{conc} presents
some concluding remarks.

\section{Doped antiferromagnets in the continuum limit}
\label{deriv}

The spin lattice of interest is defined by (\ref{dag}), but with $J$ replaced
by a position dependent exchange integral, $J_{ij}$, so
\beq
\frac{d\sv{ij}}{d\tau}=-J_{ij}\sv_{ij}\times(\sv_{i+1,j}+\sv_{i-1,j}+
\sv_{i,j+1}+\sv_{i,j-1}).
\eeq
Following \cite{kompap}, to obtain a well-defined continuum limit for this 
system we, think of it as a lattice of dimers, that is, spin pairs.
First, partition the square lattice chess-board fashion into black and
white sublattices, with site $(i,j)$ white (black) if $i-j$ is even (odd).
Associate to each white site $(i,j)$ the dimer consisting of $(\sv_{ij},
\sv_{i,j+1})$. The white sites can be labelled by a pair of (unconstrained)
integers $(\alpha,\beta)=((i+j)/2,(i-j)/2)$. We then call the two spins of
the dimer $\av_{\alpha\beta}=\sv_{ij}$ and $\bv_{\alpha\beta}=\sv_{i,j+1}$.
Note that, while $\av$ lives on the white lattice and $\bv$ on the black,
both sets of variables are labelled with reference to the white lattice.
Their evolution is governed by
\bea\label{aspin}
\frac{d\av_{\alpha\beta}}{d\tau}&=&
-J_{\alpha-\beta,\alpha+\beta}(\bv_{\alpha,\beta-1}+\bv_{\alpha\beta}+
\bv_{\alpha-1,\beta}+\bv_{\alpha-1,\beta-1})\\
\label{bspin}
\frac{d\bv_{\alpha\beta}}{d\tau}&=&
-J_{\alpha-\beta,\alpha+\beta+1}(\av_{\alpha+1,\beta}+\av_{\alpha+1,\beta+1}+
\av_{\alpha,\beta+1}+\av_{\alpha,\beta}).
\eea

We now think of the original $(i,j)$ lattice as having spacing $\eps>0$,
small, so that the $(\alpha,\beta)$ lattice has spacing $\delta=
\sqrt{2}\eps$, and assume that $\av_{\alpha\beta}$, $\bv_{\alpha\beta}$ and
$J_{ij}$ have well defined continuum limits $\avc(\eta,\xi)$, 
$\bvc(\eta,\xi)$
and $\Jc(\eta,\xi)$, where $\eta=\alpha\delta$, $\xi=\beta\delta$,
so that $\av_{\alpha+1,\beta}=\avc(\eta,\xi)+\delta\avc_{\eta}+\frac{1}{2}
\delta^2\avc_{\eta\eta}+\cdots$, and so on. Note that $\Jc(\eta,\xi)=
J_{\alpha-\beta,\alpha+\beta}$, so
\beq
J_{\alpha-\beta,\alpha+\beta+1}=\Jc(\eta+\frac{1}{2}\delta,\xi+\frac{1}{2}
\delta)=\Jc(\eta,\xi)+\frac{\delta}{2}(\Jc_\eta+\Jc_\xi)+\frac{\delta^2}{8}
(\Jc_{\eta\eta}+\Jc_{\xi\xi}+2\Jc_{\eta\xi})+\cdots.
\eeq
Substituting into (\ref{aspin}), (\ref{bspin}) and discarding terms of
order $\delta^3$ or higher, one obtains
\bea\label{acon}
2s\delta\avc_t&=&-\Jc\avc\times[4\bvc-2\delta(\bvc_\eta+
\bvc_\xi)+\delta^2(\bvc_{\eta\eta}+\bvc_{\xi\xi}+\bvc_{\eta\xi})]\\
2s\delta\bvc_t&=&-\Jc\bvc\times[4\avc+2\delta(\avc_\eta+
\avc_\xi)+\delta^2(\avc_{\eta\eta}+\avc_{\xi\xi}+\avc_{\eta\xi})]\nonumber\\
\label{bcon}
&&\qquad-\bvc\times[2\delta(\Jc_\eta+\Jc_\xi)\avc+\delta^2(\Jc_\eta+
\Jc_\xi)(\avc_\eta+\avc_\xi)+\frac{\delta^2}{2}(\Jc_{\eta\eta}+\Jc_{\xi\xi}
+2\Jc_{\eta\xi})\avc],
\eea
where $t=2s\delta\tau$ is a rescaled time variable. 

These equations simplify somewhat if we introduce the auxilliary fields
\beq
\mv=\frac{1}{2s}(\avc+\bvc),\qquad
\nv=\frac{1}{2s}(\avc-\bvc).
\eeq
Note that $\mv\cdot\nv=0$ and $|\mv|^2+|\nv|^2=1$.
Since we expect neighbouring spins to almost anti-align, $\mv$
should be small, so we will assume, consistently
with (\ref{acon}), (\ref{bcon}), that $|\mv|=O(\delta)$, whence
$|\nv|=1+O(\delta^2)$.  Then, again
neglecting terms of order higher than $\delta^2$, one finds
\bea\label{m1}
\mv_t&=&-(\cd_\eta+\cd_\xi)[\Jc\, 
\mv\times\nv]+\frac{\delta}{4}[2\Jc\, \nv\times
(\nv_{\eta\eta}+\nv_{\xi\xi}+\nv_{\eta\xi})+(\Jc_\eta+\Jc_\xi)\nv\times
(\nv_\eta+\nv_\xi)]\\
\label{n1}
\delta\, \nv_t&=&4\Jc\, \mv\times\nv-\delta\Jc\, \nv\times(\nv_\eta+\nv_\xi)-
\delta(\Jc_\eta+\Jc_\xi)\mv\times\nv-\frac{\delta^2}{4}(\Jc_\eta+\Jc_\xi)
\nv\times(\nv_\eta+\nv_\xi).
\eea
Equation (\ref{n1}) may be solved for $\mv$,
\beq
\label{m2}
\mv=\frac{\delta}{4}\big[\frac{1}{\Jc}\nv\times\nv_t-\nv_\eta-\nv_\xi\big]
+O(\delta^2),
\eeq
which may then be substituted into equation (\ref{m1}). One finds that
all the terms involving derivatives of $\Jc$ miraculously cancel, and
\beq
\nv\times\nv_{tt}=\Jc^2\nv\times(\nv_{\eta\eta}+\nv_{\xi\xi})+O(\delta).
\eeq
To leading order, therefore, $\nv$ is governed by the $O(3)$ sigma model
equation on $\R\times\R^2$ with metric $dt^2-(d\eta^2+d\xi^2)/\Jc(\eta,
\xi)^2$,
which is (\ref{**}) up to relabelling.
It is slightly more natural, however, to define continuum position
variables $x=i\eps$, $y=j\eps$ aligned with the original spin lattice, 
rather than $\eta$, $\xi$ aligned with the white sublattice.
Since the $x,y$ coordinates are obtained from  $\eta,\xi$
by a rotation (through $45^\circ$), 
$\cd_\eta^2+\cd_\xi^2=\cd_x^2+\cd_y^2$, so we again obtain
(\ref{**}), but with $J(x,y)=J_{ij}$. To reconstruct the actual spin
dynamics, given a solution of (\ref{**}), we must choose $\eps>0$ small,
construct the auxilliary field $\mv$ from (\ref{m2}),
\beq
\mv=\frac{\eps}{2\sqrt{2}}
\big[\frac{1}{J}\nv\times\nv_t-\nv_y\big],
\eeq
extract $\avc$ and $\bvc$, then sample these on the white and
black sublattices, respectively, of the lattice of spacing $\eps$. 

The above derivation is closely modelled on that of Komineas and 
Papanicolaou \cite{kompap}, so we have omitted some of the details of
the calculation. The main difference is that we have included the
extra terms due to the position dependence of $\Jc$. In the end, all these
extra terms cancel, to leading order, and we are left with the continuum
PDE one would have first guessed on naive grounds. This is somewhat
surprising, and it would be satisfying to have a more direct argument for
it than the rather complicated derivation given here.

\section{The geodesic approximation}
\label{geoapp}

Equation (\ref{**})  is the variational equation for the action 
\beq
S=\frac{1}{2}\int dt\, dx\, dy\,
\sqrt{|\eta|}\, \cd_\mu\nv\cdot\cd_\nu\nv\, \eta^{\mu\nu}
=\int dt\, (T-V) 
\eeq
where
\beq
T=\frac{1}{2}\int dx\, dy 
\frac{1}{J(x,y)^2}\left|\frac{\cd\nv}{\cd t}\right|^2,\qquad
V=\frac{1}{2}\int dx\, dy
\left(\left|\frac{\cd\nv}{\cd x}\right|^2+
\left|\frac{\cd\nv}{\cd y}\right|^2\right)
\eeq
are identifed as the kinetic and potential energy functionals respectively.
Note that while $T$ depends on $J$, $V$ does not. Hence Belavin's and
Polyakov's analysis of the {\em static} homogeneous
 model \cite{belpol} carries
over unchanged to the doped system. There is a topological lower
energy bound
\beq
\label{belpolbou}
V[\nv]\geq 4\pi|m|,
\eeq
$m$ being the topological degree of the
map $\nv:\R^2\cup\{\infty\}\ra S^2$. Let us define complex coordinates 
$z=x+iy$ on the spatial plane and $u=(n_1+in_2)/(1-n_3)$ on the
target sphere (the latter being the image of $\nv$ under stereographic
projection from $(0,0,1)$ to the equatorial plane). Then the bound 
(\ref{belpolbou}) for $m\geq 0$ is attained if and only
if $u(z)$ is a rational map of algebraic degree $m$. 
Since such rational maps minimize $V$ (globally within their homotopy class,
in fact) they are automatically stable static solutions of the model.
If we choose the
boundary value of $\nv$ at spatial infinity to
be $(0,0,1)$, the general degree 1 bubble is
\beq\label{u0}
u(z)=\chi^{-1}e^{-i\psi}(z-a)=:u_0(z;\chi,\psi,a),
\eeq
$\chi\in(0,\infty)$, $\psi\in[0,2\pi]$ and $a=a_1+ia_2\in\C$ being constants 
interpreted as the bubble's width, internal phase and position respectively.

Following Ward \cite{war} and Leese \cite{lee}, we will study the low energy
dynamics of a single bubble within the geodesic approximation of Manton
\cite{man}. The idea is that $\M_1$, the four dimensional space 
 of static bubbles, parametrized by $\chi,\psi,a_1,a_2$, which we will
denote collectively as $q^i$, is a flat valley 
bottom in the space of all degree 1 maps $\R^2\ra S^2$, 
on which $V$ attains its
minimum value of $4\pi$. If a bubble is given a small amount
of kinetic energy, its subsequent motion is confined close to $\M_1$ by
conservation of energy $E=T+V$, suggesting that a collective coordinate
approximation wherein the motion is constrained to $\M_1$ for all time is 
sensible. So we assume that
\beq\label{colco}
u(z,t)=\chi(t)e^{-i\psi(t)}(z-a(t)),
\eeq
substitute (\ref{colco}) into $S$, and obtain variational equations for
$\chi,\psi,a$. Note that $V\equiv 4\pi$, constant, for all fields of the
form (\ref{colco}), and $T$ is quadratic in time derivatives, 
so the action reduces to
\beq
S=\frac{1}{2}\int dt\, g_{ij}\dot{q}^i\dot{q}^j,\qquad
g_{ij}(q):=\int\frac{dx dy}{J(x,y)^2}
\frac{4}{(1+|u_0(z;q)|^2)^2}\frac{\cd u_0}{\cd q^i}
\frac{\cd\bar{u}_0}{\cd q^j},
\eeq
which is the action for geodesic motion on $\M_1$ with respect to the
metric $g=g_{ij}(q)dq^idq^j$. The conceptual framework and validity of this
approximation are discussed at length in \cite{mansut}.

The metric $g$, and hence the dynamics, depends
strongly on $J(x,y)$. If $J$ is constant, we are studying the standard
sigma model, and it is found that $\chi$ and $\psi$ are frozen by infinite
inertia ($g_{\chi\chi}=g_{\psi\psi}=\infty$) 
and $g=4\pi J^{-2} da\, d\bar{a}$, so bubbles
just travel at constant velocity \cite{war}.  At the other extreme,
if $J(x,y)=1+x^2+y^2$, we effectively have the sigma model on a round
two-sphere, and the bubble dynamics is much richer \cite{spe}. This choice
has $J$ unbounded, which is presumably unphysical, however. If we
assume that $J$ remains bounded, then $\chi$ and $\psi$ are frozen by
an essentially identical argument to Ward's. Without loss of 
generality, we can set $\psi=0$. The width $\chi$ remains a free, but frozen,
parameter. The metric on $\M_1^\chi$,
 the width $\chi$, phase 0 leaf of $\M_1$, is then
\beq
g=f_\chi(a)(da_1^2+da_2^2),
\eeq
where
\beq\label{fgeom}
f_\chi(a)=
\int\frac{dz\, d\bar{z}}{J(z,\bar{z})^2}\, 
\frac{4\chi^{-2}}{(1+\chi^{-2}|z-a|^2)^2}
=\int \frac{4\,  du\, d\bar{u}}{(1+|u|^2)^2}\, \frac{1}{J(\chi u+a)^2},
\eeq
the integral of $J(\chi u +a)^{-2}$ over the standard unit sphere, $u$ being
a stereographic coordinate. If $J$ is sufficiently simple, 
$f_\chi(a)$ can be computed explicitly. Note that
\beq
\lim_{\chi\ra 0} g=\frac{4\pi}{J(a)^2}(da_1^2+da_2^2), 
\eeq
which is the bubble rest mass, $M=4\pi$,
 times the inhomogeneous metric on the
physical plane. In the limit of vanishing width, therefore, bubble
trajectories tend to geodesics, not just in $\M_1^\chi$, but also in the
physical plane. One may regard $g$ as a smeared out version, due to the
bubble's finite core size, of the metric on physical space.

\section{Domain walls}
\label{domwal}

\begin{figure}[htb]
\centering
\includegraphics[scale=0.5]{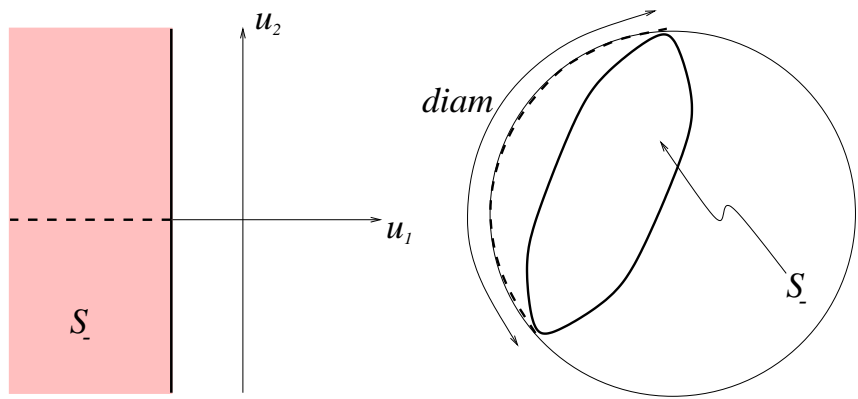}
\caption{\sf The region $S_-\subset S^2$ defined in the
derivation of $f_\chi$, equation (\ref{f}), right, and its stereographic 
image, left. The dashed curve is the great circular arc whose length is
the metric space diameter of $S_-$.
}
\label{figsphere}
\end{figure}

The simplest and most experimentally feasible type of inhomogeneous doping is
the domain wall. Let us assume that $J$ depends only on $x$, is constant
outside a small neighbourhood of $x=0$ and rises monotonically from
$J_-$ to $J_+$ as $x$ traverses this neighbourhood. Such a $J(x)$ would arise
from doping one end of an antiferromagnet but not the other.
If the domain wall is very narrow, we may idealize it by a step function,
\beq
J(x)=\left\{\begin{array}{ll} J_+ & x\geq 0\\
J_- & x<0.\end{array}\right.
\eeq
Then from (\ref{fgeom}) we see that $f_\chi$ depends only on $a_1$ and
\beq\label{f0}
f_\chi(a_1)=\frac{A_-}{J_-^2}+\frac{4\pi-A_-}{J_+^2}
\eeq
where $A_-$ is the area of $S_-\subset S^2$, the region on which 
${\rm Re}(\chi u+a)<0$. The stereographic image of $S_-$ is the half-plane to
the left of the straight line $u_1=-a_1/\chi$. The corresponding 
boundary curve
$\cd S_-\subset S^2$ is a circle whose diameter (in the metric space sense)
is the length of any great circular arc lying in $S_-$, intersecting 
$\cd S_-$ orthogonally. One such arc has stereographic image $\{t+i0\in\C\, :
\, t\in(-\infty,-a_1/\chi]\}$. The geometry of the situation is depicted in 
figure \ref{figsphere}.
Hence
\beq\label{diam}
{\rm diam}(S_-)=\int_{-\infty}^{a_1/\chi}dt\frac{2}{1+t^2}=\pi-
2\tan^{-1}\frac{a_1}{\chi}.
\eeq
To compute the area of any disk in $S^2$ with
diameter ${\rm diam}$, we can use
spherical polar coordinates with the ``North pole'' placed at the
centre of the disk. Then
\beq\label{Adiam}
A({\rm diam})=\int_0^{2\pi}d\varphi\int_0^{\frac{1}{2}{\rm diam}} 
d\theta\, \sin\theta
=2\pi\big(1-\cos\frac{1}{2}{\rm diam}\big).
\eeq
Together, (\ref{f0}), (\ref{diam}) and (\ref{Adiam}) yield
\beq\label{f}
f_\chi(a)=\frac{2\pi}{J_-^2}\left[1-\frac{a_1}{\sqrt{\chi^2+a_1^2}}\right]
+\frac{2\pi}{J_+^2}\left[1+\frac{a_1}{\sqrt{\chi^2+a_1^2}}\right].
\eeq
Note that, despite the discontinuity in $J$, $f_\chi$
is smooth, and that $f_\chi$
is a decreasing function of $a_1$ with $\lim_{a_1\ra\pm\infty}f_\chi(a_1)
=4\pi J_{\pm}^{-2}$. If $J$ is smoothed around $x=0$, these properties, on
which we will base all our dynamical arguments,
persist, although the exact formula
(\ref{f}) is lost. 

\begin{figure}[htb]
\centering
\includegraphics[scale=0.5]{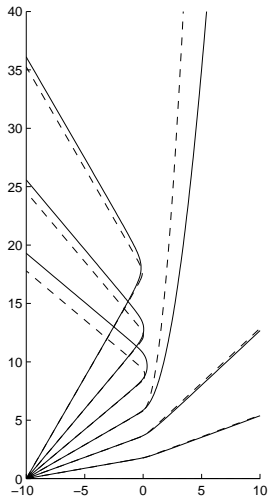}
\caption{\sf Scattering of bubbles of width $\chi=0.5$ (solid curves) and
width $\chi=0.1$ (dashed curves) off a domain wall, 
in the geodesic approximation.
The region to the left of the vertical line $x=0$ has $J=J_-=1$, while that
to the right has $J=J_+=2$. Note that bubbles are reflected if their 
incidence angle exceeds $30^\circ$, in accordance with Snell's law.
}
\label{fig1}
\end{figure}

To analyze geodesic flow in $\M_1^\chi$, we note that
the Lagrangian governing $a(t)$ is 
$
L=\frac{1}{2}f_\chi(a_1)(\dot{a}_1^2+\dot{a}_2^2)$. 
The time evolution conserves both energy $e=L$ and
the momentum conjugate to $a_2$, $p_2=f_\chi(a_1)\dot{a}_2$. Since
geodesic trajectories are independent of initial speed, we may set $E\equiv
1$. Then
\beq\label{econ}
1=\frac{1}{2}f_\chi(a_1)\dot{a}_1^2+\frac{p_2^2}{2f_\chi(a_1)}
\eeq
for all time. Consider a bubble fired obliquely at the domain wall from the
left, that is, with $a_1(0)<0$, $\dot{a}_1(0)>0$, $\dot{a}_2(0)\neq 0$.
Since $f_\chi(a_1)$ is decreasing, $a_1(t)$ must have a turning point
if $|p_2|$ (hence $|\dot{a}_2(0)|$) is too large, else (\ref{econ}) is
violated. Hence a bubble which hits the wall with two large an incidence 
angle is reflected from it, just like a light ray undergoing total internal
reflection. In fact, the analogy with refraction is much deeper. Let us
assume that the incident bubble has velocity $v_-(\cos\theta_-,\sin\theta_-)$
as $t\ra -\infty$, $-\pi/2<\theta_-<\pi/2$, so that it impinges on the wall,
and is transmitted with velocity $v_+(\cos\theta_+,\sin\theta_+)$
as $t\ra \infty$. Then by conservation of $e$ and $p_2$,
\beq
\frac{2\pi}{J_+^2}v_+^2=\frac{2\pi}{J_-^2}v_-^2 \qquad\mbox{and}\qquad
\frac{4\pi}{J_+^2}v_+\sin\theta_+=\frac{4\pi}{J_+^2}v_-\sin\theta_-
\eeq
whence one sees that
\beq\label{S}
\frac{1}{J_+}\sin\theta_+=\frac{1}{J_-}\sin\theta_-.
\eeq
This is precisely Snell's law of refraction, where we identify $J_-^{-1}$,
$J_+^{-1}$ as the refractive indices of the materials on either side of the 
wall. Hence, total internal reflection occurs if
\beq
|\theta_-|>\sin^{-1}(J_-/J_+).
\eeq
Note that (\ref{S}) is independent of $\chi$, as it must be, since the
sharp domain wall $J(x)$ is invariant under dilation of the physical
plane $(x,y)\mapsto (\lambda x,\lambda y)$.  
It is straightforward to solve the 
geodesic equations for $g=f_\chi(a_1)(da_1^2+da_2^2)$ numerically and 
confirm the behaviour predicted. Figure \ref{fig1} depicts bubble scattering
off a sharp domain wall with $J_+/J_-=2$ at a range of incidence angles.

Although we have derived Snell's law within the framework of the geodesic
approximation, one can rederive it using only field theoretic conservation
laws. By Noether's theorem and $t$ and $y$ translation invariance, the
full field theory conserves  $E=T+V$ and
\beq
P_2=\int dx\, dy\,
\frac{1}{J(x)^2}\frac{\cd\nv}{\cd t}\cdot\frac{\cd\nv}{\cd y}.
\eeq
The key point is that the effective spacetime metric $\eta$ is, except in
a narrow strip containing the $y$-axis, the Minkowski metric, but with
the ``speed of light'' differing on either side of the wall. Given a static
solution strongly localized in one of the two flat regions, therefore,
we can produce (almost) exact moving solutions by an appropriate Lorentz
boost. Let $\nv_0(x,y)$ be the static bubble given by $u(z)=\chi^{-1} z$. 
The boosted
bubble with velocity $v_-(\cos\theta_-,\sin\theta_-)$,
\beq
\nv_-(t,x,y)=\nv_0(\gamma_-(x\cos\theta_--y\sin\theta_--v_-t),
x\sin\theta_-+y\cos\theta_-),
\eeq
where the contraction factor is $\gamma_-=(1-v_-^2/J_-^2)^{-\frac{1}{2}}$,
tends to an exact solution as $t\ra-\infty$. Using this as initial data
at $t=-\infty$, let us consider the field's subsequent evolution. Presumably
the bubble impinges on the domain wall. Let us assume that the bubble is
transmitted, and that its interaction with the wall is elastic, so no energy
is lost to radiation. Then, as $t\ra \infty$ it is once again a free bubble
moving at constant velocity, so must approach some translate of
\beq
\nv_+(t,x,y)=\nv_0(\gamma_+(x\cos\theta_+-y\sin\theta_+-v_+t),
x\sin\theta_++y\cos\theta_+),
\eeq
with contraction factor $\gamma_+=(1-v_+^2/J_-^2)^{-\frac{1}{2}}$. It is
straightforward to compute $E$ and $P_2$ for $\nv_\pm$ as $t\ra\pm\infty$, 
yielding
the conservation laws
\beq
M\gamma_+=M\gamma_-\quad\mbox{and}\quad
\frac{v_+\gamma_+\sin\theta_+}{J_+^2}M=
\frac{v_-\gamma_-\sin\theta_-}{J_-^2}M,
\eeq
where $M=4\pi$ is again the rest energy of the bubble,
which together imply (\ref{S}). Note that $M$ is independent of $\chi$, so
this argument holds even if the interaction with the wall somehow changes
the bubble's width. The limiting factor on the applicability of (\ref{S}) is
thus the assumption of elastic scattering off the wall, which one expects
to hold at small incidence velocities, rather than the 
validity of the geodesic approximation
itself.

\section{Circular lenses}
\label{cirlen}

It is interesting to consider the interaction of bubbles with curved
domain walls. The observation that bubbles undergo refraction, with
$J^{-1}$ identified as the refractive index, immediately gives a simple
``geometric optics'' model of their trajectories, which should hold whenever
the bubble width is large relative to the domain wall width, but small
relative to its radius of curvature. This raises the interesting
possibility that lenses can be
constructed which could be used to focus bubbles onto a target area. 
Such a lens would consist of a thin region bounded by a pair of
circular arcs of large radius, within which $J$ is suppressed relative to
its value outside the region. This is a rather complicated choice of
$J(x,y)$ for the purposes of computing $g$, so we shall here test the
feasibility of antiferromagnetic lenses more indirectly, by studying the
scattering of bubbles off complete disks of suppressed $J$, for which 
$g$ can again be computed explicitly. Clearly,
such circular ``lenses'' are far from the thin lenses used for focussing
in optics. Nevertheless, by comparing bubble trajectories incident on a
circular lens with their geometric optics analogues, we can provide 
strong evidence that bubble focussing is possible.

We shall again make a thin domain wall approximation, so that $J(z)$
is piecewise constant, taking one value on some disk and another value
elsewhere. By conformal invariance we can, without loss of generality, assume
the disk has unit radius. Hence
\beq
J(z)=\left\{\begin{array}{cc}
J_-& |z|< 1\\
J_+& |z|\geq1\end{array}\right.
\eeq
with $J_-<J_+$. We can again use formula (\ref{f0}), but now $S_-\subset
S^2$ is the region on which $|\chi u+a|<1$. This is again a disk, whose
stereographic image is bounded by the circle of radius $\chi^{-1}$ centred
at $-a/\chi$. By rotating around the $n_3$ axis, this disk can be centred
on the positive $u_2$ axis, so that the segment of this axis from
$u_2=(|a|-1)/\chi$ to $u_2=(|a|+1)/\chi$ becomes a great circular arc
in $S_-$ intersecting $\cd S_-$ orthogonally. Hence
\beq
{\rm diam}(S_-)=\int_{(|a|-1)/\chi}^{(|a|+1)/\chi}dt\, \frac{2}{1+t^2}=
2\big(\tan^{-1}\frac{|a|+1}{\chi}-\tan^{-1}\frac{|a|-1}{\chi}\big),
\eeq
which, together with (\ref{Adiam}) and (\ref{f0}) gives
\beq
f_\chi(a)=\frac{2\pi}{J_-^2}\left[1-\frac{\chi^2+|a|^2-1}{\sqrt{\chi^4+
2\chi^2(|a|^2+1)+(|a|^2-1)^2}}\right]+
\frac{2\pi}{J_+^2}\left[1+\frac{\chi^2+|a|^2-1}{\sqrt{\chi^4+
2\chi^2(|a|^2+1)+(|a|^2-1)^2}}\right].
\eeq
Note that $f_\chi$ depends on $|a|$ alone, is smooth, and has
$\lim_{|a|\ra\infty}f_\chi(a)=4\pi J_+^{-2}$, as one would expect. The
dependence on the bubble width $\chi$ cannot, as in the case of a straight
sharp domain wall, be scaled away. Unless $\chi \ll 1$, therefore, we
should expect bubble trajectories to diverge significantly from there
optical analogues. 

\begin{figure}[htb]
\centering
\includegraphics[scale=0.5]{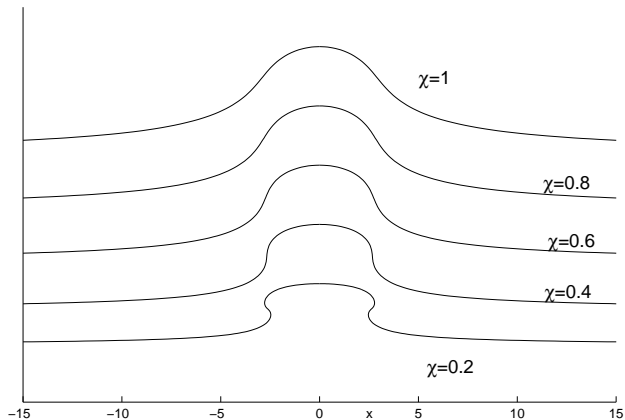}
\caption{\sf Generating curves for $\M_1^\chi$ isometrically embedded as
a surface of revolution in $\R^3$ for various $\chi$, in the case
of a sharp disk inhomegeneity with $J_-=1$, $J_+=2$. The surfaces are
swept out as these curves are rotated about the axis $x=0$. Note
that a neck develops as $\chi$ becomes small.}
\label{figdisk1}
\label{figsor}
\end{figure}

For $\chi$ sufficiently large $\chi > \chi_1$ ($\approx 0.17$ if
$J_+/J_-$=2), $(\M_1^\chi,g)$
may be isometrically embedded as a surface of revolution in $\R^3$, using
the method described in \cite{mcgspe}. Figure \ref{figsor} shows 
numerically generated generating curves for these surfaces of revolution for
various values of $\chi$: the surfaces themselves are swept out as the 
curves are rotated about an axis vertical in the page. It is not surprising
that the isometric embedding fails for $\chi$ small. Recall that in the
limit $\chi\ra 0$, $g$ tends to a multiple of the metric on the physical
plane, which for this choice of $J$ is clearly singular. It can be roughly
visualized as the flat plane with the unit disk cut out, expanded by a
linear factor of $J_+/J_-$, its edge then re-attached to the hole. The
surfaces depicted in figure \ref{figsor} are on the way to this geometry
as $\chi$ gets small. An interesting point to note is that a pinched
neck develops in the surface for $\chi<\chi_2$ ($\approx 0.36$
for $J_+/J_-=2$). The two
circles swept out by the points where the tangent to the curve is vertical
are both geodesics. Hence there is a bifurcation in the geodesic flow
at $\chi=\chi_2$, a pair of periodic geodesics appearing as $\chi$ is
reduced through $\chi_2$. Although we have deduced the existence of these
periodic geodesics visually, it is easy to show that they exist for all
$\chi<\chi_2$ (even if $\chi<\chi_1$, so that no isometric embedding exists)
by appealing to conservation of energy $e$ and angular momentum
$p_\theta=f_\chi(|a|)|a|^2\dot{\theta}$, where $a=|a|e^{i\theta}$. 
The corresponding bubble trajectories are circular orbits centred on 
$z=0$, one inside the disk and one just outside. Such trajectories have
no optical analogue. 

\begin{figure}[htb]
\centering
\begin{tabular}{ccc}
(a)&(b)&(c)\\
\includegraphics[scale=0.3]{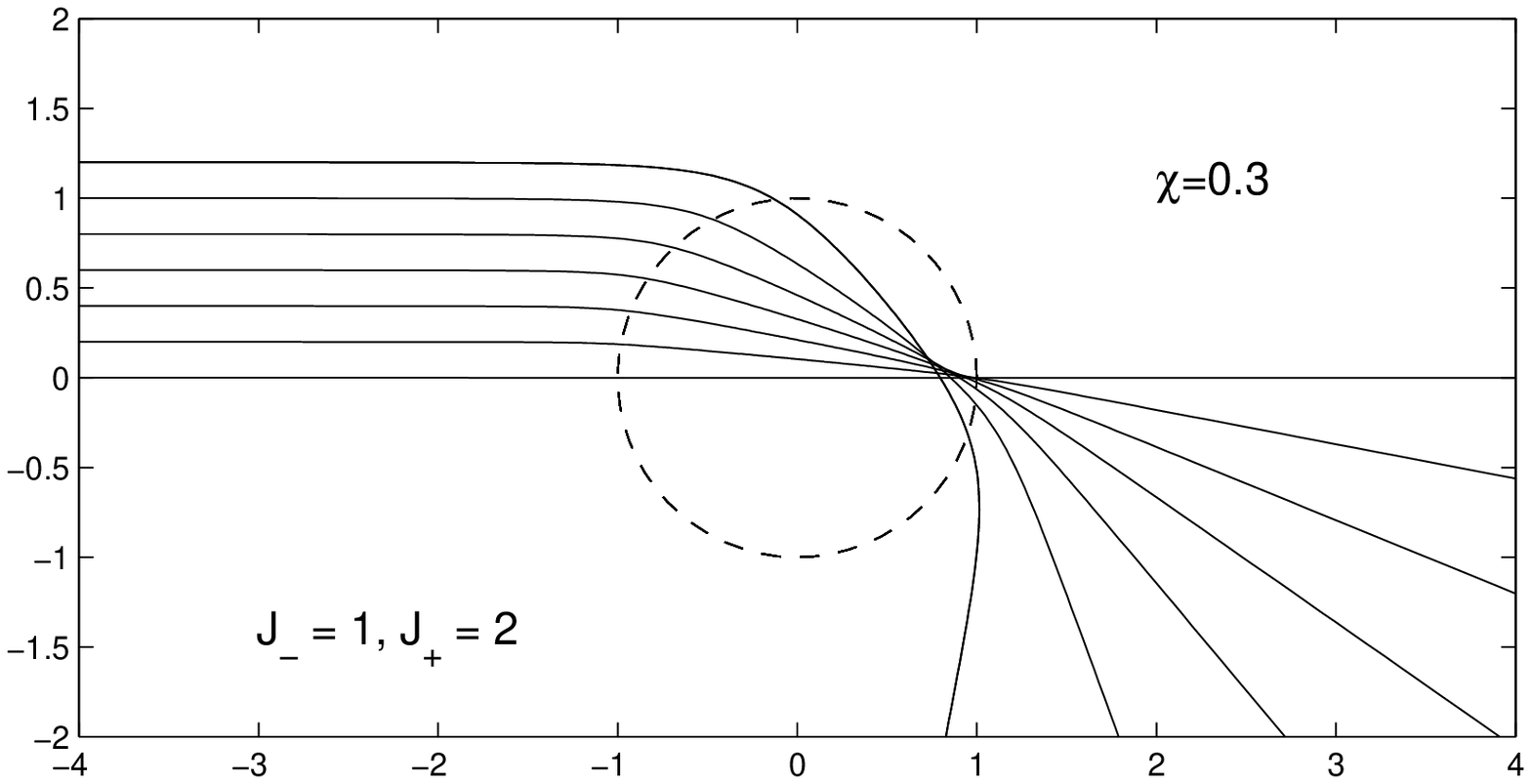}&
\includegraphics[scale=0.3]{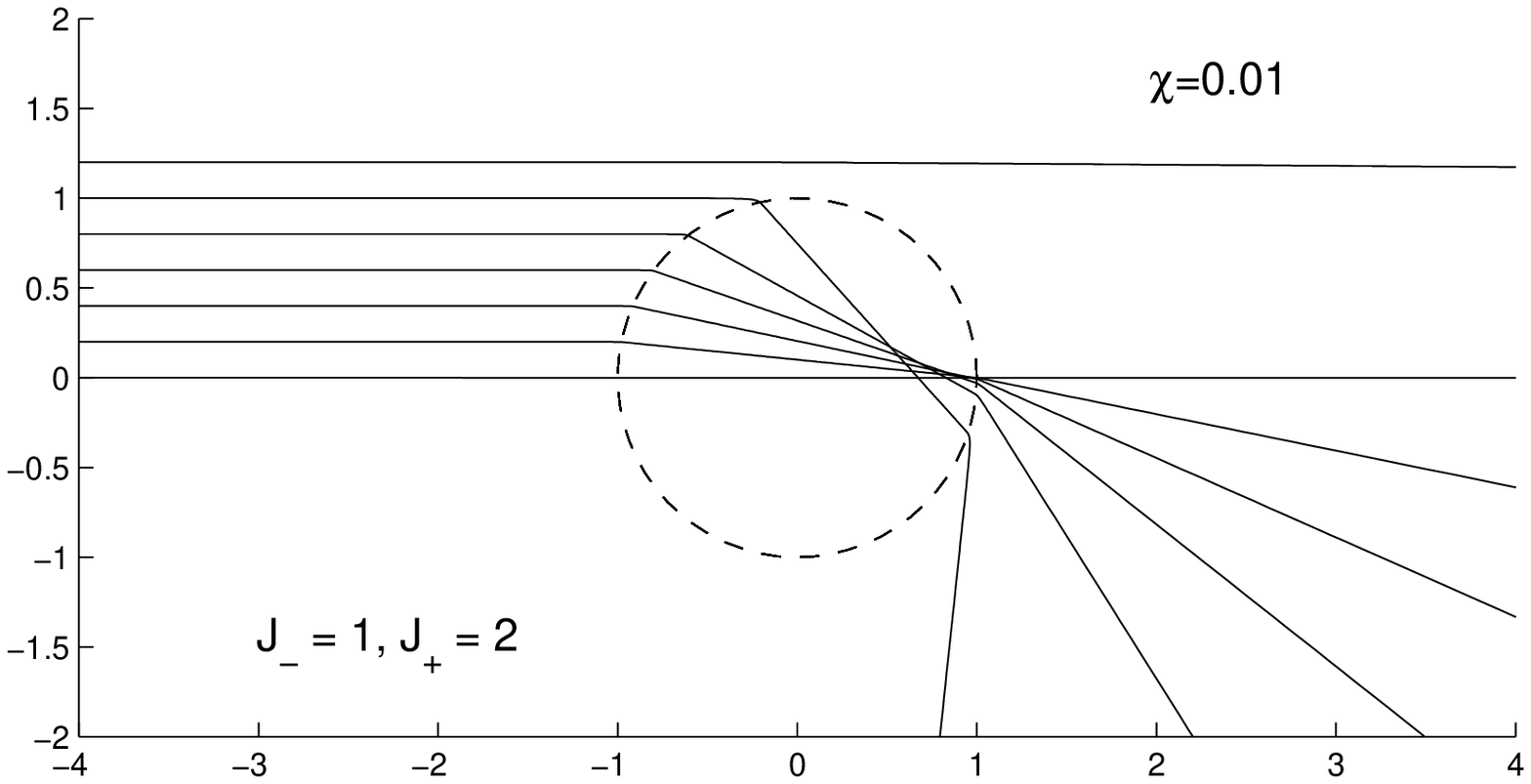}&
\includegraphics[scale=0.3]{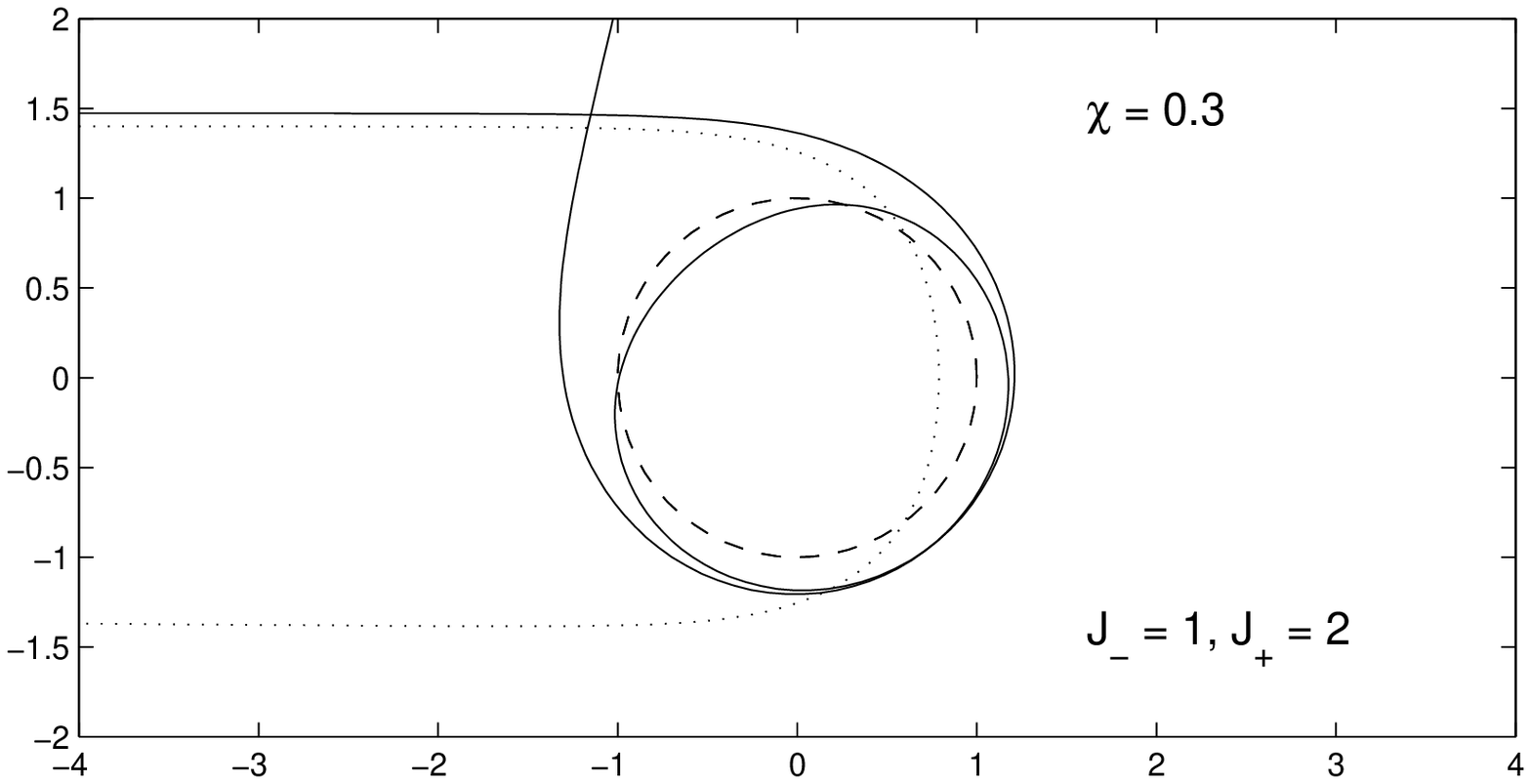}\\
\end{tabular}
\caption{\sf Scattering of bubbles off a circular ``lens'' with $J_+/J_-=2$:
(a) and (b) show the expected focussing effect for $\chi=0.3$ and $0.01$
respectively, while (c) shows the sensitive dependence of the trajectory
on its initial data when it winds around the neck of $\M_1^\chi$ ($\chi=0.3$
again).
}
\label{figdisk2}
\end{figure}

It is again straightforward to solve the geodesic problem numerically.
Figure \ref{figdisk1} shows the scattering of bubbles initially moving
along parallel trajectories for two different choices of $\chi$. A focussing
effect is apparent, and the trajectories look optical in the case of very 
narrow bubbles. To make a quantitive comparison, we 
present in figure \ref{figdisk2} plots of the exit angle $\theta$ of the
bubble as a function of its impact parameter $b=\lim_{t\ra-\infty}a_2(t)$
for a variety of widths $\chi$,
in comparison with the exit angle for light rays,
\beq
\theta_{\rm light}(b)=2\left(\sin^{-1}\frac{J_-}{J_+}b-\sin^{-1}b\right).
\eeq
Clearly, bubble scattering is increasingly optical as they become narrow,
as one expects. Even for moderately narrow bubbles the geometric optics
picture works well except for very glancing impacts ($b$ close to $1$).
One area where the geometric optics model fails badly is scattering with
$b$ just exceeding $1$. Light rays in this case miss the disk completely,
so $\theta_{\rm light}(b)\equiv 0$. Bubbles, on the other hand, can become
almost trapped by the neck in $\M_1^\chi$, and scatter with nontrivial
$\theta$, having wound several times around the disk. The dependence of
$\theta$ on $b$ is very sensitive in this regime, see figure \ref{figdisk2}.

\begin{figure}[htb]
\centering
\includegraphics[scale=0.5]{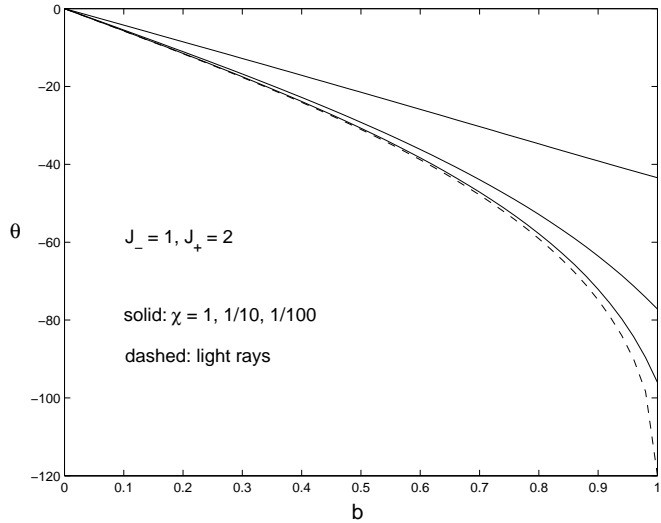}
\caption{\sf 
The exit angle of a bubble trajectory scattering off a circular lens
with $J_+/J_-=2$, as a function of impact parameter
for $\chi=1$, $0.1$, $0.01$ (solid curves, top to bottom), in comparison
with the exit angle for light rays (dashed).
}
\label{figdisk3}
\end{figure}

If we choose $J_->J_+$, the disk causes bubble trajectories to diverge rather
than focus, as one would expect. Again the trajectories are increasingly
optical as $\chi$ becomes small. A significant difference is that $\M_1^\chi$
can never be isometrically embedded in $\R^3$, and there are
no periodic geodesics.

\section{Bubble guides}
\label{bubgui}

\begin{figure}
\centering
\begin{tabular}{cc}
(a)& (b)\\
\includegraphics[scale=0.4]{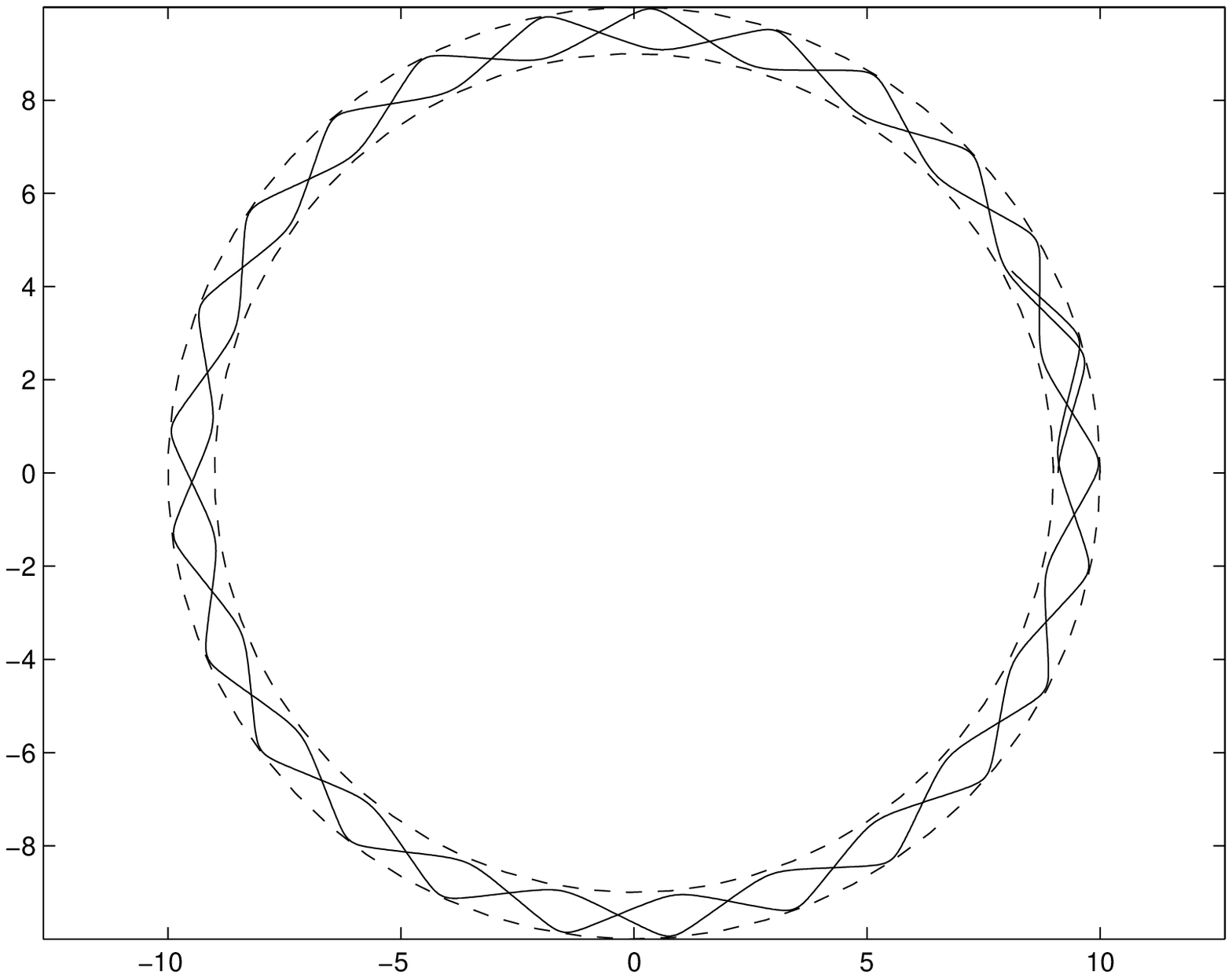}&
\includegraphics[scale=0.4]{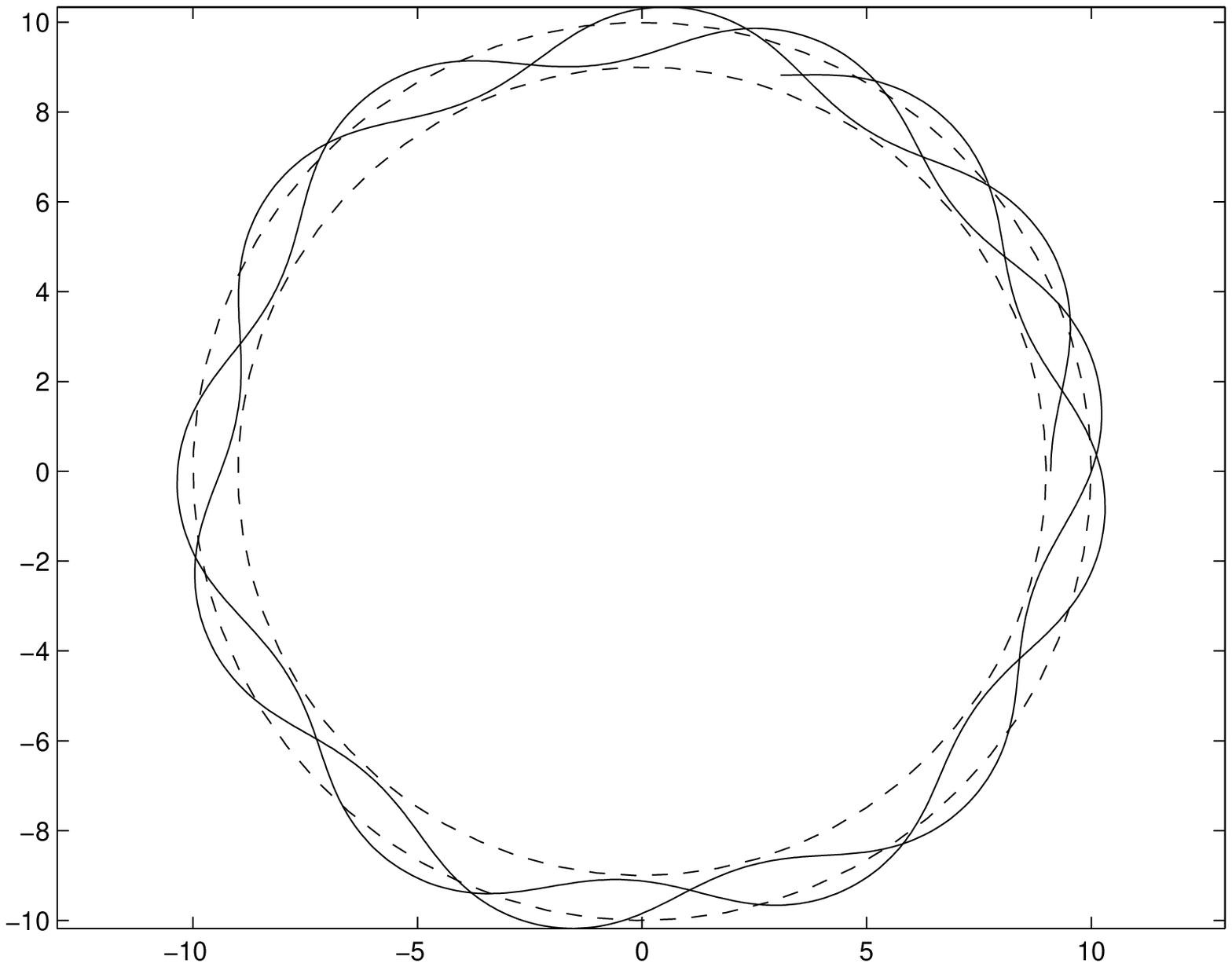}\\
\end{tabular}
\caption{\sf A circular bubble guide, analogous to a fibre optic cable. The
region between the dashed circles has $J$ suppressed by a factor of 2.\,
Total internal reflexion keeps narrow bubbles ($\chi=0.1$)
between the walls, (a).
Even a bubble as wide as the track itself ($\chi=1$)
is guided along it, (b).
}
\label{fig2}
\end{figure}

 The
total internal reflexion phenomenon might be used to construct the analogue
for bubbles
of fibre optic cables: doped tracks of low $J$ in a high $J$ substrate
which would guide bubbles along a prescribed path. A simple example, 
for which $f_\chi(a)$ can still be obtained exactly, is an annular track 
with
\beq
J(z)=\left\{\begin{array}{cc}
J_-& r_1<|z|<r_2\\
J_+& \mbox{elsewhere}\end{array}\right.
\eeq
where $J_-<J_+$ as before. Once again (\ref{f0}) holds, but now $S_-$
(or rather its stereographic image) is the annular region centred on
$-a/\chi$ bounded by radii $r_1/\chi$ and $r_2/\chi$. The area of this
region can be computed in similar fashion to the previous cases (by finding
the diameters of its inner and outer disks), yielding the formula
\beq
f_\chi(a)=\frac{4\pi}{J_+^2}+2\pi\big(\frac{1}{J_+^2}-\frac{1}{J_-^2}\big)
[h(|a|,\chi,r_1)-h(|a|,\chi,r_2)],
\eeq
where
\beq
h(\alpha,\chi,r)=\frac{\chi^2+\alpha^2-r^2}{\sqrt{\chi^4+2\chi^2(\alpha^2
+r^2)+(\alpha^2-r^2)^2}}.
\eeq
Bubble trajectories within such an annular bubble guide with $J_+/J_-=2$
and $r_1=9$, $r_2=10$, obtained numerically within the geodesic
approximation, are depicted in figure \ref{fig2}. 
A narrow bubble ($\chi=0.1$, compared with the track width of
unity) initially moving within the annulus is trapped by the domain
walls and guided around it. Even a bubble as wide as the track
($\chi=1$) is guided along it, although the bubble centre now strays 
beyond the domain walls, and the trajectory looks far from optical.

\section{Conclusion}
\label{conc}

We have argued that, in the near continuum regime, a two-dimensional
isotropic Heisenberg antiferromagnet with inhomogeneous doping should
be well modelled by the relativistic $O(3)$ sigma model on a spatially
inhomogeneous spacetime. By conformal invariance, such a model supports
static Belavin-Polyakov lumps, which we call bubbles. The trajectories
of these bubbles are predicted to behave like light rays propagating
in an inhomogeneous medium, with $J^{-1}$, the inverse exchange integral,
identified with the refractive 
index. In particular, bubbles incident on a straight domain wall are
refracted in accordance with Snell's law, and total internal reflexion can
occur. This result was derived both within the framework of the geodesic
approximation, and by appeal to field theoretic conservation laws. 
Its validity is not contingent on the validity of the geodesic approximation
per se, but rather on the assumptions that the interaction of
bubble and domain wall is elastic, which should certainly be true for low
impact speeds, and that the scattering does not cause the bubble to collapse
immediately (it must survive long enough to travel far from the
domain wall for our conservation law argument to apply). Bubble collapse
is an intense focus of current research, and it is still unclear
whether collapse is generic. 

In the present context, it would be perverse
to try to settle this issue by lattice simulation of the sigma model, since
this is itself merely an approximation to the more fundamental spin lattice
(\ref{dag}). Numerical simulations of (\ref{dag}) seeking the refraction
phenomena noted here are the logical next step, but lie beyond the scope
of this paper. Even if bubble collapse does turn out to be a serious
problem, there is still hope that bubble refraction may occur in real-life
systems because, just as in the case of ferromagnets, higher order
interactions may contribute extra terms which stabilize the bubble
\cite{kirpok}. These
terms destroy the conformal invariance of the sigma model, so there would be
corrections to Snell's law due to the $J$ dependence of the bubble's rest
mass, but the basic refractive behaviour may persist. The main obstacle to
seeing such refraction in a real antiferromagnet is that, since neighbouring
spins almost anti-align, bubbles have almost zero magnetization, and are
thus difficult to observe experimentally. 

To illustrate the possibility, it is instructive to consider the spatially
inhomogeneous system studied recently by Piette {\it et al} \cite{piezakbra},
\beq
S=\int dt\, dx\, dy\, \frac{1}{2}\cd_\mu\nv\cdot\cd^\mu\nv+
\frac{1}{4}[(\cd_\mu\nv\cdot\cd_\nu\nv)(\cd^\mu\nv\cdot\cd^\nu\nv)-
(\cd_\mu\nv\cdot\cd^\mu\nv)^2]-\frac{\mu(x,y)}{2}(1-n_3^2),
\eeq
which may be loosely interpreted as an antiferromagnet with homogeneous
and isotropic exchange integral ($J\equiv 1$), but with inhomogeneous
on-site easy-axis anisotropy (the potential coefficient $\mu(x,y)$ depends on
position). This system supports approximate bubble type solutions,
localized in regions of constant $\mu$, but their
width is no longer a free parameter, and both their width and
rest-mass now depend on position. Let us assume again that we have
domain wall doping, so $\mu(x,y)=\mu_-$ for $x<0$, $\mu(x,y)=\mu_+>\mu_-$
for $x\geq 0$. Then we expect the rest mass of a bubble in the right
half-plane, $M_+$, to exceed that of a bubble in the left half-plane,
$M_-$ (Piette {\it et al}\, find $M_+/M_-\approx 1.029$ for $\mu_-=0.6$,
$\mu_+=0.7$, for example). What happens when a bubble moving with speed
$v_-$ at incident angle $\theta_-$ hits the domain wall? By energy 
conservation, it must be reflected if $\gamma_-M_- <M_+$, independent of
$\theta_-$. Assume $v_-$ exceeds this threshold,
$v_-^2>1-M_-^2/M_+^2$, and the bubble crosses
the barrier, escaping with speed $v_+$ at angle $\theta_+$. Then energy
conservation $M_-\gamma_-=M_+\gamma_+$ fixes $v_+$ as a function of
$v_-$, and clearly $v_+<v_-$. Again, the $y$-component of linear momentum
is conserved, so $v_-\gamma_-\sin\theta_-M_-=v_+\gamma_+\sin\theta_+ M_+$,
which together with energy conservation imply
\beq
v_+\sin\theta_+=v_-\sin\theta_-.
\eeq
Since $v_->v_+$ refraction does occur (the bubble bends away from the
normal), and the bubble must undergo total internal reflexion if
$|\theta_-|$ exceeds the $v_-$-dependent Brewster angle 
$\sin^{-1}[v_+(v_-)/v_-]$. 

Turning to theoretical high energy physics, the $O(3)$ sigma model has long 
been a favoured model for those interested in soliton dynamics in 
geometrically exotic spacetimes \cite{spe,spetor,spestr,comgib,rom,covzak}.
Our observation raises the possibility that some exotic spacetimes may
be artificially engineered in the laboratory. Furthermore, since
Snell's law emerges from conformal invariance (the soliton's rest mass is 
independent of $J$) and very general
conservation law arguments, soliton refraction
should not be a special feature of this particular
model. One would expect it to be a feature of Yang-Mills
instanton dynamics in
$4+1$ dimensions, for example. 

\section*{Acknowledgements}

The author thanks Stavros Komineas for helpful correspondence
about antiferromagnets.

\end{document}